\documentclass[final,3p,times,authoryear]{elsarticle}

\usepackage{amssymb}
\usepackage{hyperref}
\hypersetup{colorlinks=true,citecolor=blue}
\usepackage{graphicx,amsmath}
\usepackage{ifthen}
\usepackage{multirow}
\usepackage[mathscr]{euscript}
\def\beq{\begin{equation}}
\def\eeq{\end{equation}}
\def\bea{\begin{eqnarray}}
\def\eea{\end{eqnarray}}
\def\nn{\nonumber}

\def\ibid{(\emph{ibid.})}

\def\fnn#1#2{\def#1{\footnote{#2}\,}}
  
\biboptions{round,comma,sort&compress}

\journal{.}

\begin{document}
\begin{frontmatter}

\title{Fashion, fads and the popularity of choices: micro-foundations for diffusion consumer theory}

\author{Jean-Fran\c{c}ois Mercure }
\ead{jm801@cam.ac.uk}
\address{Department of Environmental Science, Radboud University, PO Box 9010, 6500 GL Nijmegen, The Netherlands}
\address{Cambridge Centre for Environment, Energy and Natural Resource Governance (C-EENRG), University of Cambridge, 19 Silver Street, Cambridge, CB3 1EP, United Kingdom}

\begin{abstract}

Knowledge acquisition by consumers is a key process in the diffusion of innovations. However, in standard theories of the representative agent, agents do not learn and innovations are adopted instantaneously. Here, we show that in a discrete choice model where utility-maximising agents with heterogenous preferences learn about products through peers, their stock of knowledge on products becomes heterogenous, fads and fashions arise, and transitivity in aggregate preferences is lost. Non-equilibrium path-dependent dynamics emerge, the representative agent exhibits behavioural rules different than individual agents, and aggregate utility cannot be optimised. Instead, an evolutionary theory of product innovation and diffusion emerges. 

\end{abstract}


\end{frontmatter}


\section{Introduction}

\fnn\fnAA{See for example \cite{Grubler1999, Nakicenovic1986, Sharif1976, Fisher1971, Mansfield1961, Farrell1993, Kwasnicki1996, Marchetti1978}.}

\fnn\fnAB{Named after \cite{Lotka1925} and \cite{Volterra1939}, normally applied to ecological systems. Volterra researched a solution to the problem of changes in fish populations in the Adriatic sea after the first World War, where the latter prevented fishermen from fishing, thus changing the balance of population growth and decline. The same Lotka-Volterra equation system had been used to model the evolution of technology, see for example \cite{Morris2003, Karmeshu1985, Farrell1993,Lakka2012,Mercure2015}.}

\fnn\fnABB{Agents interact with one another through the economy only, i.e. through prices clearing markets, but agents do not influence the behaviour of other agents.}

\fnn\fnAC{\cite{Young2009} positions diffusion as stemming from contagion, social influence, social learning, or a mixture of these. Here we do not specifically mean credibility as information asymmetry \citep{Akerlof1970}, although it could easily be included in the model.} 

\fnn\fnAD{Ultimately, we must acknowledge that agents are not infinitely lived, and are born without any knowledge at all. Every consumption habit that people learn will have been demonstrated by someone somewhere, in most likelihood someone near, e.g. a family member or friends.}

\fnn\fnAE{We define `objective' here as homogenous knowledge of goods held by all agents over which they have clear preferences, thus statistically rankable.}

\fnn\fnAF{Stochastic theories of economic behaviour take the same stance and mathematical form \citep{Aoki2007}}

\fnn\fnAG{\cite{Brock2001} analyse \emph{choices} (e.g. voting) but not \emph{adoption} (e.g. purchasing), the latter involving a rate of change of market shares and thus time.}

\fnn\fnAI{See \cite{Kreindler2014, Matejka2015, Young2009, Kreindler2013}.}

The diffusion of innovations is not an instantaneous process, nor is it an equilibrium process. In 1943, researchers studied the uptake of farming innovations by USA farmers, and observed a particular structure in which neighbours contributed to their acceptance, leading to a bell-shaped curve in rates of adoption \citep{Ryan1943}. In 1963, Rogers published his seminal work on the diffusion of innovations, in which he observed a recurrent feature: the uptake of innovations followed a slow exponential take-off, followed by a fast and steady linear adoption profile, which eventually slowed down in a saturation process \citep{Rogers2010}. This stems from interactions between adopters, starting with early adopters who influence the choices of various categories of later adopters. This process is now famously known as $S$-shaped diffusion profiles. 

It is clear that diffusion processes are sufficiently slow that they can be well observed under a standard frequency of economic measurement (e.g. quarterly, yearly). This represents a process of change in which adoptions of innovations or new products influence and promote further adoptions of the same, and intrinsically implies a process out of equilibrium. For example, people having seen a movie influence others whether or not to go and see the same film, and thus the number of sales is never at a steady state, never in equilibrium. Even when external (exogenous) variables do not change (e.g. income and prices), sales of products grow and decline over time. Many examples of studies of S-shaped diffusion processes exist in the diffusion lliterature\fnAA. Just as in ecology, in which the growth of populations can take place at the expense of others, it has been argued that diffusion profiles follow Lotka-Volterra types of ecological equations\fnAB. In general, the diffusion process is one that involves communication across complex networks, some kind of epidemic process. Mathematically, diffusion problems are non-linear and therefore, standard equilibrium consumer theory, using Constant Elasticity Substitution (CES) functions may not be appropriate to study for the substitution of products when their diffusion is not stable. 

We face a case of process involving inter-agent interactions. Trends and fashions are important factors driving the diffusion of innovations, in which, typically, the diffusion of a product reinforces its own ability to diffuse \citep{Rogers2010}. Such crowd effects arise only in models that include multi-agent interactions. Complexity theory \citep{Anderson1972} contends that the addition of multi-agent interactions in a model leads to the emergence of new structures, which depend predominantly on the nature of the interactions, and to a lesser degree so on the rules of behaviour of agents in isolation \citep{Arthur2014, Anderson1989, Kirman2011}. With interactions, what maximises the utility of individual agents does not necessarily maximise aggregate utility \citep{Kirman1992}. Crowd effects have been studied in contexts such as financial markets \citep{Arthur1997}, technological lock-ins \citep{Arthur1989}, information cascades and herd behaviour in firm decisions \citep{Bikhchandani1992,Banerjee1992}, the diffusion of innovations \citep{Rogers2010,Young2009, Kreindler2014}, and social influence in discrete choices \citep{Brock2001, Durlauf2010}. In that context, since the representative agent is a theory without explicit multi-agent interactions,\fnABB and that it assumes that a steady state equilibrium takes place, it does not model the diffusion of innovations very well.

A deeper study of the problem reveals that information transfer between agents and its credibility is in large parts at the root of the issue \citep[see e.g.][]{Young2009}.\fnAC In standard consumer theory, agents have full knowledge of all goods in all markets and possess innate ordered transitive preferences over all of these things. However, if one allows that agents are not born with complete `objective' knowledge (albeit with preferences) of all goods in all markets, and of all new goods being introduced, agents must learn in order to consume.\fnAD Information acquisition models are known to behave similarly to disease propagation models \citep[][]{Arthur1993, Lane1997, Young2009}, as they possess the typical structure of an evolving complex network. The adoption of innovations extensively follows the same dynamics \citep[e.g. as known in empirical work, ][]{Mansfield1961, Fisher1971}. Information acquisition leads to hierarchical structures based on trust, from which emerge information cascades \citep{Bikhchandani1992,Banerjee1992}. If knowledge of all goods is not equally shared by all agents, then knowledge of goods in markets, and their value, ceases to be `objective', as different bits of knowledge become shared only by subsets of the population (social groups), and no product can be unambiguously established across the population as superior, on average, to any other.

The question is, therefore: how do we credibly introduce information acquisition in standard models of consumer behaviour, and what does it mean for quantitative economic modelling, consumer tax policy, market placement and innovation policy? More generally, the question can be framed as, what are the rules governing the behaviour of the representative consumer in a theory where multi-agent interactions are included? The representative agent may still be a utility maximiser, however he unavoidably will have \emph{additional behavioural traits} that individual agents do not have: the crowd effects. These emergent crowd effects, as is known in complexity theory, cannot be exhaustively enumerated. The macro-theory may then differ considerably from the micro-theory. Considering this, then, we may ask ourselves what the relationship is between a theory of learning agents, if there is one, with standard classical consumer theory.

The diversity of agents is known to strongly influence rates of adoption of innovations \citep{Rogers2010}. The inclusion of agent preference heterogeneity in consumer theory lies in the realm of Discrete Choice Theory (DCT) \citep{Domencich1975, McFadden1973}. \cite{Anderson1992} demonstrate how DCT models are equivalent to standard substitution models under budget constraints, and pave a way to define the representative consumer from statistical micro-foundations. This average man displays elasticities representative of the aggregate response to economic contexts of an underlying population of heterogeneous agents, behaving stochastically and/or having different preference ranking orders.\fnAF Even in cases of imperfect information, optimal choices of information seeking agents facing costs of information acquisition, although influenced by prior beliefs, match the standard multinomial logit \citep{Matejka2015}. DCT thus establishes clear micro-foundations for deriving the aggregate behaviour of a population of non-interacting heterogeneous agents in equilibrium, in which the representative agent has the same behavioural rules as individual utility maximisers in isolation. The issue with DCT is that it is a static picture, the list of choices cannot change, because any change in the list of products available can lead to irreversible radical `flips' in market allocations which violate the equilibrium assumption.

In a real economy, agents have incomplete knowledge of markets and are subject to the influence of other agents in their choices, attracted to novelty products. Schumpeterian entrepreneurs continuously strive to introduce new differentiated products in attempts to secure profits by capturing the interest of consumers. When following trends, agents inevitably make choices that violate the ranking of preferences that would have been theirs without interactions with their peers while learning about products, existing or new. Thus the number of agents who know any product at any time continuously changes, as trends come and go, making knowledge `subjective' (i.e. context-dependent).

In this article, we derive the simplest possible theoretical micro-economic foundations for a non-equilibrium consumer theory involving the diffusion of innovations, and the introduction of differentiated products by entrepreneurs, and show how such a model can improve standard Constant Elasticity of Substitution (CES) models. Following the steps of \cite{Brock2001}, we modify a standard discrete choice model to include social interactions. However, instead of looking for equilibria \citep[as in ][]{Brock2001}, we instead focus on the dynamics \citep[as in][]{Arthur1993}. Indeed, the connection between discrete choices and the dynamics of diffusion is not clearly established.\fnAI The reason to focus on dynamics is that fads and fashions, originating from social influence, inherently change over time, and this is critical for consumer markets. 

For example, imagine the case of the market for moving pictures. People may, on average, see one film or less per week. People often choose which films to watch through peer recommendations. If agents require, say, between 1 and 3 recommendations to decide to watch a film, and see one film a week, the timescale of diffusion for a film to take off will be of several weeks, longer than the quarterly timescale of economic measurement. This ordinary system is not in equilibrium under the measurement timescale. In fact, for an equilibrium to exist in a market when prices and income are constant, choices must settle to a steady state under a timescale much shorter than the measurement frequency, which is often not the case for consumer goods and services (especially durables), and the analysis of social influence in equilibrium of \cite{Brock2001}, use for example in choices between schools or other long term social choices, does not apply to the problem here.\fnAG As for moving pictures, most consumer goods markets are in continuous flux and change, and thus, we explicitly choose not to look for fixed points in the theory, and time as a variable becomes important.

For the sake of clarity of our model, we first review standard discrete choice consumer theory (section~\ref{sect:II}), which we subsequently modify. We build from micro-foundations a theory of interacting consumers by incorporating social influence (i.e. learning) in a discrete choice model, and explain the non-equilibrium dynamical implications (sections~\ref{sect:III}). Were there no new products appearing in the market, consumption habits would indeed converge towards a small number of homogenous goods in equilibrium. This never happens in reality, and thus we explicitly include innovation, and the model leads to product diversity with a composition that changes over time. We show that by tuning the strength of multi-agent interactions to zero, standard consumer theory is restored, where a phase transition takes place. We discuss the roles of innovation, diffusion and the entrepreneur, of an evolutionary nature (section~\ref{sect:V}). Finally (section~\ref{sect:VI}), we conclude that in a model of heterogenous learning agents, the meaning of facts and knowledge is subjective to context, and thus the concept of utility of the representative agent, normally used in welfare theory, is ambiguous in the present model. We note that while consumer tax policy acts as incentive orienting the diffusion trajectory, problems of optimal tax policy have no solution.

\section{Review of equilibrium consumer theory} \label{sect:II}
\subsection{Choice modelling in equilibrium}

\fnn\fnBA{This distribution is appropriate because it represents the statistical distribution of extreme values}

\fnn\fnBB{Here, when $U_i >> U_j, P_{ij} = 1$, while when $U_j >> U_i, P_{ij} = 0$, with fractional values between. The utility difference required for either option to have $1/(1+e)$ the probability of the other is $|U_i - U_j| = \sigma$, the width of the distributions.}

\fnn\fnBC{We use the $<>$ brackets to express an average, e.g. $\left< X \right> = \sum_i P_i X_i$.}

We review standard DCT in equilibrium, upon which we later build a non-equilibrium DCT. DCT provides a consistent methodology to determine aggregate choice of a group of non-interacting heterogenous agents. Consumer theory is not always expressed in the form of DCT, but as reviewed by \cite{Anderson1992}, all equilibrium consumer theories are equivalent. We consider agents who make choices between products to purchase, in diversified markets, in order to maximise their utility. Agents do not substitute variable quantities of all products, not due to budget constraints, but because they cannot or do not need to (e.g. choosing between cars, restaurant meals). Product diversity means that goods exist in sets that achieve the same purpose, but that possess varying characteristics and prices.  

Heterogeneous utility maximisers have preferences that we describe using probability distributions, for which the Gumbel distribution of width $\sigma$ \citep[or double exponential, ][]{Domencich1975} is appropriate:\fnBA 
\beq
P(U > U_i) = F_i(U) = e^{-e^{-\left({U - U_i \over \sigma}\right)}}, \quad f_i = e^{-e^{-\left({U - U_i \over \sigma}\right)}}e^{-\left({U - U_i \over \sigma}\right)}dU,
\eeq
where $f$ is the probability distribution, $F$ the cumulative distribution, $U$ is an arbitrary utility value, $U_i$ is the mean utility associated to  product $i$ and $\sigma$ is the diversity of the population. We thus describe the choice process as the comparison between the frequency distribution of utility maxima. We define a Linear Random Utility Model \citep[LRUM,][]{Anderson1992} in terms of a number of socio-economic variables that make agents heterogenous, for choice options $i$
\beq
U_i^* = \beta^1_i V^1_i + \beta^2_i V^2_i + \beta^3_iV^3_i + ... + \epsilon_i,
\label{eq:randUt}
\eeq
where the $\beta$s represent various socio-economic parameters (elasticities), the $V$s are the matching socio-economic variables (e.g. income, prices, preferences, etc), and $\epsilon_i$ is an error term. The star denotes the stochastic nature of $U_i^*$ (it is distributed). In the binary case, we compute the frequency at which option $i$ generates, on average over a diverse population, more utility than option $j$. This yields a convolution of the cumulative distribution ($F_j$) of one option with the frequency distribution ($f_i$) of the other, where for an arbitrary $U$:
\bea
P_{ij} \left( U_j^*>U \big| U = U_i^* \right) &=& \int_{-\infty}^{\infty} f_i(U-U_i) F_j(U-U_j) dU \nn\\
& = & {1 \over 1 + \exp(-{U_i-U_j \over \sigma})},
\label{eq:binary}
\eea
yielding the well-known logistic function of the binary logit model (BL).\fnBB

To treat more than two options simultaneously, we evaluate the probability that the arbitrary utility value $U$ given by one option is higher than all possible choices in a set of $n$ possibilities,
\beq
P \left( U >  \max \left[U_1^*, U_2^*, ... U_n^* \right] \right) = P \left( U >   U_1^* \right) P \left( U >   U_2^* \right) ... P \left( U >   U_n^* \right),
\label{eq:PUg}
\eeq
and then determine what is the probability that the utility of option $i$ is greater or equal to $U$, given that $U$ is greater than the utility of all other choices:
\bea
P\left (U_i^*>U \big| U =  \max \left[U_1^*, ... U_{i-1}^*, U_{i+1}^*, ... U_n^* \right]\right)\nn\\
= \int_{-\infty}^{\infty} f_i(U-U_j) \overline{F}(U-\overline{U}) dU,
\eea
where $\overline{F}(U-\overline{U})$ is the cumulative utility distribution of the `representative consumer' with unknown mean $\overline{U}$. This is similar to the binary logit case. The representative consumer has a distribution of preferences which is the product of all distributions, from which the utility of the representative consumer $\overline{U}$ can be isolated: 
\bea
\overline{F}(U - \overline{U}) &=& \exp \left( -\sum_j e^{-\left({U-U_j \over \sigma}\right)} \right) = e^{-e^{-\left({U- \overline{U} \over \sigma}\right)} }, \nn\\
\overline{U} &=& \sigma \log \left( \sum_j e^{U_j \over \sigma} \right)
\eea
In statistical physics, this would be called the `partition function'. Replacing $U_j$ for $\overline{U}$ in eq.~\ref{eq:binary}, we obtain the standard multinomial logit (MNL),
\beq
P_i = { e^{U_i \over \sigma} \over \sum_j e^{U_j \over \sigma}}, \quad \sum_i P_i = 1.
\label{eq:MNL}
\eeq
This well behaved function tells us what the probability choice of option $i$ is in equilibrium, for a context given by the variables $V^x_y$ of the random utility in eq.~\ref{eq:randUt}. The key assumption required is that all agents know equally all options available to them and choose the one giving them the greatest utility, even if they have different preferences, and that the list of choices never changes. 

We can also obtain the MNL by taking partial derivatives of the utility of the representative consumer:
\beq
P_i = {\partial \overline{U} \over \partial U_i}\bigg|_{U_j} = { e^{U_i \over \sigma} \over \sum_j e^{U_j \over \sigma}}.
\label{eq:partial}
\eeq

All changes in prices or other variables of the individual utilities $U_i$ that increase the utility of the representative consumer correspond to Pareto improvements. In equilibrium, therefore, the utility of the representative consumer is maximised and thus constant with respect to changes in variables.

The average utility across agents is different:\fnBC
\beq
\left< U \right> = {\sum_i U_i e^{U_i \over \sigma} \over \sum_j e^{U_j \over \sigma}} = {\partial \over \partial {1 \over \sigma}} \left( {\overline{U}  \over \sigma} \right),
\eeq
We define an `entropy' type function $\mathbb{S}$ as the difference between $\overline{U}$ and $\left< U \right>$, called the \emph{consumer surplus}:
\beq
\sigma \mathbb{S} = \sigma {\partial \overline{U} \over \partial \sigma} = \sigma \log \left( \sum_j e^{U_j \over \sigma} \right) - {\sum_i U_i e^{U_i \over \sigma} \over \sum_j e^{U_j \over \sigma}} = \overline{U} - \left< U \right>.
\label{eq:entropy}
\eeq
It is effectively an entropy \citep{Anderson1988} since it can be written as
\beq
\mathbb{S} = -\sum_i P_i \log P_i, 
\eeq
maximised in equilibrium. Note that the function $\overline{U} = \mathbb{S} + \left< U \right>$ is conceptually the same as Helmholtz' free energy for a system of physical particles, which is minimised in thermodynamic equilibrium. The utility of the representative consumer has also been given interpretations in terms of information theory \citep[information minimisation, see e.g. ][]{Anas1983}, and welfare economics \citep{dePalma2009, Small1981},\footnote{Commonly called the log-sum rule.} and has been used directly as an indicator of aggregate welfare for choices in public policy \citep[see e.g.][]{McFadden2001}. This is a reversible system in which time does not appear, nor innovation or the introduction of new products. The introduction of new products would change $\mathbb{S}$ and would be irreversible, and therefore involves non-equilibrium dynamics, ultimately leading to a new equilibrium.

DCT is an aggregate model of economic behaviour for the representative agent that correctly represents the sum of heterogeneous individual behaviour that is consistent with standard consumer theory. From these detailed microeconomic first principles, the model of constant elasticity of substitution (CES) can be derived (see \ref{sect:AppA}). Maximising the utility $U$ of a CES model under budget constraint $Y$, with prices $p_i$, of the form
\beq
U(X_i) = \left( \sum_j \left[X_j\right]^\rho \right)^{1\over \rho}, \quad Y = \sum_j X_i p_i
\label{eq:CES}
\eeq
yields the MNL \citep{Anderson1992}. This indicates that the MNL is not only consistent with a standard utility maximisation model; it gives it micro-foundations. The choice of products in the MNL thus matches general equilibrium theory. CES models underpin much of Computable General Equilibrium (CGE) modelling \citep[see e.g. ][]{Dixon2013}, widely used for quantitative economic analysis.

\subsection{Limits of equilibrium consumer theory}
\fnn\fnCA{The unchanging stock of knowledge is implied because the MNL in equilibrium requires an unchanging list of choice options, as otherwise the sudden addition of an option could lead discontinuously to a radically different market allocation, and a discontinuity in the utility of the representative consumer.}
\fnn\fnCB{E.g. see \cite{Young2001}. See also \cite{Burke2011} for a good introduction.}

Equilibrium consumer models are quite restrictive: they \emph{require} that agents (1) \emph{do not interact} with one another, (2) \emph{know from birth} all products in all markets, (3) are able to express a complete ranking of their preferences for every product, and (4) have an \emph{unchanging stock of knowledge}.\fnCA Such a system is in fact equivalent to systems of individual agents interacting with the economy only through demand/supply signals, but not with each other. Unchanging stocks of knowledge imply no novelty products, which implies no innovating entrepreneurs, and no diffusion of innovations, and thus no entrepreneurial profits nor productivity change in the sense of \cite{Solow1957}.

Indeed, if one allows that agents find value in consuming products that \emph{other agents} also consume (non-price interactions), then the corresponding discrete choice theory model that incorporates fashions and trends inevitably takes a radically different form. If agent $k$ \emph{finds value $V^4_i(k,\ell)$ in consuming product $i$ that other agents $\ell$ also consume}, then a term that links the utility between agents arises \citep[see e.g.][]{Durlauf2010}:
\beq
U_i(k) = \beta^1_i V^1_i(k) + \beta^2_i V^2_i(k) + \beta^3_iV^3_i(k) + ... + \alpha f\left( \sum_{\ell} \beta^4_i V^4_i(k,\ell) \right) + \epsilon_i(k),
\label{eq:IntLRUM}
\eeq
When such interactions are introduced, the absolute ranking of options breaks down in ways unpredictable with standard consumer theory \citep[including DCT, ][]{Brock2001}, since preferences depend recursively on preferences. However, if this term is tuned to a smaller and smaller value, standard theory is restored again.\footnote{Note that an optimisation is guaranteed to have a unique solution only when $\alpha = 0$.} 

Taking a step back, it is clear that preferences are \emph{subjective}, and vary over space and time, rather than have a unique equilibrium. Agent rankings of preferences that exclude social influence could be seen as unrealistic, since what then drives agent preferences, if not his/her surroundings, cultural and family context, social class, social role, etc? \citep[e.g. see][]{Benhabib2010}. Trends and fashions \emph{do} occur, and the popularity of novelty products \emph{can} generate significant prosperity to particular entrepreneurs, while old products become disused. They also drive product diversity through  consumption that is homogenous within social groups and heterogeneous across \fnCB. 

\section{A theory for interacting consumers in economics} \label{sect:III}
\subsection{The problem of heterogeneous knowledge: when everyone knows different things}

We introduce a model in which agents, before purchasing goods, need to learn about their existence: people are born without knowledge of any product. Product use is social group dependent. Product diversity is large, and there is no need to assume that agents know of all available products at all times. Manufacturers adjust product diversity over time to match consumer tastes, in such a way that consumers do not need to search the whole market for goods, they are offered goods mostly tailored to their differentiated tastes. Firms consider that consumers do not need to be made aware of all the possible products that, individually, they are not likely to desire (firms do not spend unnecessarily on unproductive marketing). 

Product diversity in vehicle markets was characterised by \cite{MercureLam2015}, who found that it is comparatively large, and the frequency distribution of vehicle purchase is related to the income distribution, and differs by region of the world, even though in principle all vehicle models are technically available everywhere. This suggests that preferences are subjective, contextual and vary geographically.

It is known in sociology that purchase choices are often made through visual influence. In vehicle markets, it has been shown empirically that choices are made through visual influence within social groups, geographical areas, even price brackets \citep[e.g.][]{McShane2012}. Households of a particular level of income and social identity do not therefore seek to know the full breadth of goods that target other income and social identity groups, as they already know in which market they will be searching. Explanations are suggested by socio-anthropological theory and observations, for example in the `anthropology of consumption' \citep{Douglas1979}, which describes consumption as an act of context-dependent social interaction.\footnote{I.e. consumption does not happen solely for maximising the utility of lone individuals, but rather, as an act of communication of social identity between several individuals.} 

Agents are unlikely to purchase goods that they have never heard of, seen, used or seen used \citep[e.g. as observed by ][]{Rogers2010}. Instead, they are more likely to purchase goods that peers own or buy (credible information sources), such that they know what they are buying \ibid. In the present model, influence happens through \emph{interactions} between agents. For example, the more people own a particular type of vehicle, the more often it is observed by peers, and the higher the likelihood of new sales of the same. In such a model, across the choice probabilities of agents, options cannot be equally weighted. Instead, they must be weighted by their \emph{popularity}. The diffusion of innovations occurs through the diffusion of knowledge, which stems from interactions across agents. This allows a theory where knowledge of the existence of new products starts from zero.

When everyone knows different things (\emph{heterogeneous knowledge}), the stock of knowledge varies between agents, with only partial overlaps between any possible pair. When communication interactions exist between agents, the stock of knowledge of each agent changes over time. If products also change over time, the stock of knowledge of agents over products does not increase indefinitely; they can forget about older products gradually discontinued (e.g. land-line phones) and learn about new ones (e.g. mobile phones). Agents increase their (subjective) utility by purchasing amongst the products that they \emph{know}, but each agent maximises his own utility on a different set of choices (with a different maximum welfare). 

Social groups are defined loosely as groups of agents having a relatively homogenous stock of knowledge and preferences, agents linked to one another by the \emph{popularity} of particular consumption habits. This gives rise to product differentiation between, for instance, social, cultural and economic classes.\footnote{E.g. some social groups, may be choosing amongst products of a completely different nature to other social groups, for example across social classes or cultural groups. As emphasised by \cite{Douglas1979}, these groups of consumption norms are often separated by price barriers.} This gives rise to a so-called local conformity/global diversity effect \citep{Burke2011}. Such a theory provides a robust basis to define \emph{bounded rationality} in discrete choices beyond `satisficing' \citep{Simon1955}, as we see below.

\subsection{Theory for strongly interacting consumers in economics}

We define a representative consumer in a bounded-rational system of information-sharing utility maximisers. The popularity/visibility of choices can be represented by correctly weighing option choice probabilities, and given that the distributions are Gumbel, and that the product of Gumbel distributions is Gumbel, the weighting is an exponent for option $i$, denoted $S_i$.

In a random sampling of information-sharing agents, considering that the relative frequency of picking product $i$-using agents is the share of the market occupied by product $i$, then $S_i$ is the market share. Re-evaluating expression~\ref{eq:PUg} by instead multiplying on each side the probabilities calculated individually for all $N$ agents, with $N_1, N_2, ... N_n$ the numbers of agents using products $1,2,...n$, ($\sum N_i = N$) we obtain
\beq
P \left( U >  \max \left[U_1, U_2, ... U_n \right] \right)^N = P \left( U >   U_1 \right)^{N_1} P \left( U >   U_2 \right)^{N_2} ... P \left( U >   U_n \right)^{N_n},\nn
\eeq
which, using $S_i = N_i/N$, can be written as 
\beq
P \left( U >  \max \left[U_1, U_2, U_3, ... U_n \right] \right) = \prod_i P \left( U >   U_i \right)^{S_i},  
\eeq
where the $S_i$ are market shares, and $\sum_i S_i = 1$. One then finds the utility of the representative consumer, as a partition function for interacting agents:
\beq
\overline{U} = \sigma \log \left( \sum_i S_i e^{U_i \over \sigma} \right).
\label{eq:repU}
\eeq
We take partial derivatives to obtain choice probabilities as in eq.~\ref{eq:partial}. Temporarily, we propose that product shares $S_i$ are single-valued functions of utilities $U_i$, as in the MNL:
\beq
P_i = {\partial \overline{U} \over \partial U_i} = {S_i e^{ U_i \over \sigma} \over \sum_k S_k e^{ U_k \over \sigma}} + \sigma { \sum_j {\partial S_j \over \partial U_i } e^{ U_j \over \sigma} \over \sum_k S_k e^{ U_k \over \sigma }}.
\label{eq:Temporarily}
\eeq
This expression is made complex by the second term involving the change in market shares $S_i$ with respect to changes in the utilities of product categories $U_i$. Effectively, as the utilities change, the number of consumers acquiring and getting to know these products change, and their ability to communicate about them changes, and thus the stock of knowledge changes. This is a recursive problem, of which the convergence must be determined.

\subsection{The non-existence of a unique and shared equilibrium set of knowledge}

We now show that if we assume an equilibrium, we obtain an absurdity. The equilibrium is a steady state; this means that all agents who could increase their utility have done so, otherwise they would change their choices, and it would not be a steady state. This means that $S_i = P_i$, i.e. the shares of the market have converged to the choices, as is the case in the MNL. We have that
\beq
P_i = {P_i e^{ U_i \over \sigma} \over \sum_k P_k e^{ U_k \over \sigma}} + \sigma { \sum_j {\partial P_j \over \partial U_i } e^{ U_j \over \sigma} \over \sum_k P_k e^{U_k \over \sigma }}.
\label{eq:nonsense}
\eeq
We show in \ref{sect:AppB} that for a general form of $P_i$, the only solution is for one of the $P_i$ to be equal to one, and all the others ($P_{j \neq i}$) equal to zero, where all agents have converged to the same preference. This would be the case if we waited for a sufficiently long time, and would lead to multiple possible equilibria \citep[one for each option, as in][]{Brock2001}. But for shorter timescales, this breaks our assumption and observations of existing product diversity. Thus we conclude that the shares $S_i$ cannot be equal to the preferences $P_i$, and thus that we cannot have an equilibrium in this system if product diversity exists. 

\subsection{Knowledge propagates like a disease}

The adoption of innovations is never instantaneous, and this is due to the fact that consuming a product takes time: consumers have dinner once a day, purchase a new car every few years, change their mobile phone when the contract finishes, and tend to use up the 80 tea bags in a box before they purchase a new box or change brand, given a change in context (price, characteristics, etc). Thus whereas preferences $P_i$ are instantaneous (if we survey customers, they are always able to tell their opinion immediately), the shares of product use $S_i$ lag behind preferences with a time delay. 

This has far-reaching consequences. Imagining that a large price or utility change takes place all of a sudden (e.g. the introduction of a tax): it does not make the shares of product use change at the same instantaneous rate. This is because agents need to, for example, finish the box of tea bags, wear out or pay off their car, finish their mobile phone contract or decide to eat out, each of which imply a different average statistical time lag, which we denote $\tau_i$. The price or utility change in $P_i$ gives an \emph{incentive} for agents to change their choice, which pulls the $S_i$ in a particular direction at a rate $1/\tau_i$.  The actual value of $S_i$ strongly depends on what its value was at a time just before the change in price or utility took place. Thus $S_i$ may lag behind $P_i$ with an average time scale $\tau_i$, but $S_i$ could also have any value depending on its history. Meanwhile, the $P_i$ also evolve as the $S_i$ change, and `aggregate tastes' evolve with agents discovering products and telling each other. 

The key outcome is that the shares $S_i$ evolve in a direction that increases (at different rates) the utility of all agents as they substitute (or not) products at replacement time, at a rate lower or equal to $\tau_i$; meanwhile, if the context changes faster that $\tau_i$, then $S_i$ never catches up with $P_i$ and the system is never in an equilibrium or steady state. Note that $S_i$ can evolve even for constant values of $P_i$, which will happen for instance after a sudden change in $P_i$ (e.g. suddenly introducing a new constant tax). Meanwhile the $P_i$ do not change unless utilities $U_i$ change, or the popularity $S_i$ changes. 

These are hallmarks of a dynamical system, in which $S_i$ is \emph{hysteretic}, and \emph{path-dependent}, i.e. $S_i$ is a multi-valued function of $U_i$, depending on configurational history. It cannot be evaluated with a standard linear regression. In such a system, equilibria and limit cycles may exist, but are not always straightforward to identify \citep{Hofbauer1998}; we do not search for their existence, but focus on the properties of the dynamics.

\begin{figure}[t]
	\begin{center}
		\includegraphics[width=0.9\columnwidth]{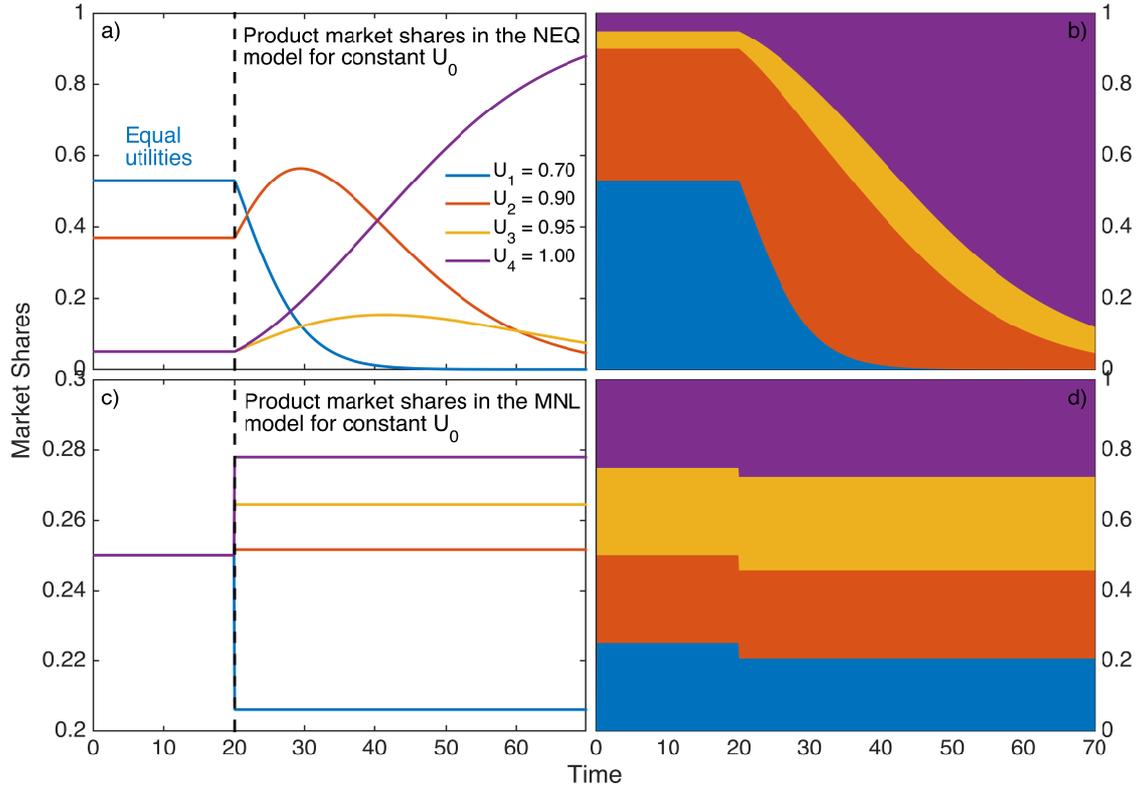}
	\end{center}
	\caption{Outcomes of the non-equilibrium (NEQ) (eq.~\ref{eq:Lotka}) and the multinomial logit (MNL) models, for a set of 4 competing products, with constant utilities $U_0$ and different shares at the starting point. At $t>20$ (in arbitrary time units), utilities becomes different. In a)-b), the NEQ model displays a gradual process of diffusion of products. In c)-d), the MNL gives constant market shares before and after $t=20$, settling instantaneously at new values.}
	\label{fig:Figure1}
\end{figure}

Thus we build on the fact that while $P_i$ provides a signal of agent preferences, it only applies to those agents who are witness to product use, and this is a fraction $S_i$ of agents. Since shares are \emph{not} actual functions of utilities, the partial derivative in the second term of eq.~\ref{eq:Temporarily} is zero. We construct the simplest possible adoption model, which involves a rate of replacement of $1/\tau_i$. In a dynamical model of adoption, there is an in-flow of shares, when people buy products, and an out-flow of shares, when people finish consuming the products. We have the preferences
\beq
P_i =  {S_i e^{ U_i \over \sigma} \over \sum_k S_k e^{ U_k \over \sigma}} 
\label{eq:Sprefs}
\eeq
The adoptions are
\beq
{dS_i^\uparrow \over dt} = {P_i \over \tau_i}{\overline{\tau} \over \tilde{\tau}}; \quad \text{with}\, {1 \over \overline{\tau}} = \sum_i {S_i \over \tau_i}; \quad {1 \over \tilde{\tau}} = \sum_i {P_i \over \tau_i},
\eeq
with $\overline{\tau},\tilde{\tau}$ the lifetime and the adoption rate, respectively. The out-flow is simply ${dS_i^\downarrow \over dt} = {S_i \over \tau_i}$, which means that the dynamical shares equation is
\beq
{dS_i \over dt} = {1 \over \tau_i}\left[ {\overline{\tau} \over \tilde{\tau}} {S_i e^{ U_i \over \sigma} \over \sum_k S_k e^{ U_k \over \sigma}} - S_i \right].
\label{eq:shares1}
\eeq
This is correct since $\sum_i {dS_i \over dt} = 0$. Furthermore, if the timescale of consumption is very short (or much shorter than the rate of observation), then we can take a limit for when $\tau_i \rightarrow 0$, using the transformation $y_i = {1 \over \tau_i} \rightarrow \infty$, multiplying \ref{eq:shares1} by $\tau_i$ on both sides:
\beq
\lim_{\tau_i \rightarrow 0} {dS_i \over dt} \tau_i = 0 = \lim_{\tau_i \rightarrow 0} {\overline{\tau} \over \tilde{\tau}} P_i - S_i = \lim_{y_i \rightarrow \infty} {\sum P_i y_i \over \sum S_i y_i} = {P_i^2 \over S_i} - S_i.
\eeq
This implies that $P_i = S_i$. Therefore, when the rate of adoption and consumption is much faster than the rate of observation, we fall back to a standard equilibrium state \citep[as in ][]{Brock2001} in which market shares equal preferences, which, if we took out the multi-agent interactions, would also lead us back to the MNL, where
\beq
S_i = {\exp\left({U_i^0 \over \sigma} \right) \over \sum_{k} \exp\left({U_k^0 \over \sigma} \right)} 
\eeq
This is the case for instance with transport decisions, which involves very short timescales, and not necessarily any social influence. 

In the case of durable goods or production capital, there may be both a decision of adoption and a decision for scrapping (e.g. machines, vehicles), therefore with a lifetime $\tau_i$ and acquisition or production rate $t_i$, which gives:
\beq
{d S_i \over dt} = S_i \left( \mathcal{F}_i - \overline{\mathcal{F}} \right), \quad \mathcal{F}_i =   {{1\over \tau_i} S_i e^{ U_i \over \sigma} \over \sum_k {1\over \tau_k}S_k e^{ U_k \over \sigma}} - {{1\over t_i} S_i e^{ U_i \over \sigma} \over \sum_k {1\over t_k} S_k e^{ U_k \over \sigma}},
\label{eq:replicator}
\eeq
where $\mathcal{F}$ is a comparative product `fitness', or competitiveness, in the market, in the evolutionary sense. This is the general form of the replicator equation used in evolutionary game theory \citep[][]{Hofbauer1998}. It can be transformed into a pair-wise exchange of market shares (imitation dynamics), a form of the Lotka-Volterra equation of population dynamics in competing species (see ~\ref{sect:AppC} for a derivation; see also \citealt{Mercure2015}), 
\beq
{d S_i \over dt} = \sum_j S_i S_j \left( A_{ij}F_{ij} - A_{ji}F_{ji} \right), \quad F_{ij} = {1 \over 1+ e^{{U_i-U_j}\over \sigma}}
\label{eq:Lotka}
\eeq
with matrices $A_{ij}$ and $F_{ij}$ the rates of substitution and pair-wise expected preferences (a binary logit), respectively. Both models have the same properties, but the latter is more convenient computationally and has been used in various studies \citep{Mercure2012, Mercure2014, Mercure2018}.

Figure~\ref{fig:Figure1} shows a comparison of model behaviour between this non-equilibrium (NEQ) replicator dynamics model and the MNL model. While the NEQ model shows continuous evolution even when utilities are constant, the MNL displays changes only when the utilities change. This means that, for example, a new constant tax will kick-off the diffusion of new products in the NEQ model, while in the MNL a tax must continuously change in order for the diffusion of products to proceed.

\subsection{The `thermodynamics' of the diffusion system}

In the present model, agents make choices that gradually take them towards ever higher utility as they learn of, and adopt new products. Do these substitutions correspond to Pareto improvements, and does this lead to equilibrium when Pareto improvements have been exhausted, even after a time longer than the characteristic time scale of the market? To determine this, we look at whether the consumer surplus (i.e. the entropy) changes with product choices. We note that these mathematics are similar in spirit to those of thermodynamics \citep{Reif1965}. We summarise three equations for the aggregate quantities: the consumer surplus, the average utility and the utility of the representative consumer, which are connected by $\mathbb{S} = \overline{U} - \left< U \right>$:

\beq
\sigma \mathbb{S} = -\sigma \sum_i P_i \log P_i = \sigma \log \left( \sum_j S_j e^{U_i \over \sigma} \right) - {\sum_i U_i S_i e^{ U_i \over \sigma} \over \sum_k S_k e^{ U_k \over \sigma}};\nn
\eeq
\beq
\left< U \right> = {\sum_i U_i S_i e^{U_i \over \sigma} \over \sum_k S_k e^{U_k \over \sigma}}; \quad \overline{U} = \sigma \log \left( \sum_i S_i e^{U_i\over \sigma} \right); \quad \text{with} \quad P_i = {S_i e^{U_i \over \sigma} \over \sum_k S_k e^{U_k \over \sigma}}.
\eeq
We can establish a number of relationships here. First, some useful equations:
\beq
{\partial P_i \over \partial S_i}\bigg|_{U_i} = {P_i \over S_i} - {P_i^2 \over S_i}; \quad \text{and} \quad
{\partial P_i \over \partial U_i}\bigg|_{S_i} = {P_i \over \sigma} - {P_i^2 \over \sigma}
\eeq
and therefore a first identity is
\beq
{\partial P_i \over \partial U_i}\bigg|_{S_i} = -{S_i \over \sigma}{\partial P_i \over \partial U_i}\bigg|_{S_i}.
\eeq

We take derivatives of the aggregate quantities:

\begin{tabular}{ r l r l }
&&&\\
${\partial \mathbb{S} \over \partial S_i}\big|_{U_i}$& = ${P_i \over \sigma S_i}\left( \sigma + \left< U \right> - U_i \right);$ & ${\partial \mathbb{S} \over \partial U_i}\big|_{S_i}$& = ${P_i \over \sigma^2} \left( \left< U \right> - U_i \right)$\\
${\partial \left< U \right> \over \partial S_i}\big|_{U_i}$& = $-{P_i \over \sigma S_i^2}\left( \left< U \right> - U_i \right);$ & ${\partial \left< U \right> \over \partial U_i}\big|_{S_i}$& = ${P_i \over \sigma} \left( \sigma - \left< U \right> + U_i \right)$\\
${\partial \overline{U} \over \partial S_i}\big|_{U_i}$& = $P_i;$ & ${\partial \overline{U} \over \partial U_i}\big|_{S_i}$& = ${\sigma P_i \over S_i}$\\
&&&\\
\end{tabular}

We see that these equations are mostly functions of the difference between the utility of one option and the average utility. This implies three `thermodyamics' identities:

\begin{tabular}{ r l }
&\\
${\partial \mathbb{S} \over \partial S_i}\big|_{U_i}$& = $-{1\over S_i} {\partial \left< U \right> \over \partial U_i}\big|_{S_i} + {2 P_i \over S_i};$ \\
${\partial \mathbb{S} \over \partial U_i}\big|_{S_i}$& = $-{S_i\over \sigma} {\partial \left< U \right> \over \partial S_i}\big|_{U_i}$ \\
${\partial \overline{U} \over \partial U_i}\big|_{S_i}$& = ${S_i\over \sigma} {\partial \overline{U} \over \partial S_i}\big|_{U_i}$ \\
&\\
\end{tabular}

Note that while the total changes in consumer surplus for changes in utility are zero, or $\sum_i {\partial \mathbb{S} \over \partial U_i}\big|_{S_i} = 0$, the changes in consumer surplus for changes in shares are
\beq
\sum_i {\partial \mathbb{S} \over \partial S_i}\big|_{U_i} = \left< {1\over S_i} \right> - {1 \over \sigma}\left< {U_i \over S_i} \right> + \left< U \right> \left< {1\over S_i} \right>
\eeq
Thus, the consumer surplus becomes large when the shares of some options are small (minimum product diversity), and is minimal when shares are all the inverse of the number of options (maximum product diversity). This means that the representative agent doesn't `like' product diversity; instead, everyone would enjoy higher utility (even if they don't know it) from using the same `best', dominant product, unless even better products are created (partly because, by construction, agents gain utility for following the choices of others). Furthermore, the total change in the utility of the representative agent for changes in individual utilities is $\left< {1 \over S_i} \right>$, and thus follows a similar rule.

\subsection{On stochasticity, irreversibility, path-dependence and lock-ins}

A system in equilibrium would be so independently of the configuration of product use and knowledge $S_i$. In the present case it is not, and furthermore, we can show that it is generally irreversible, meaning that the state of the market will depend on configurational history. The simplest way to demonstrate irreversibility is to evaluate the cross derivatives of the consumer surplus, i.e. the entropy \citep[see e.g.][p. 157]{Richmond2013}:
\beq
{\partial \mathbb{S} \over \partial U_i \partial S_i} = \left({P_i \over \sigma^2 S_i} - {2P_i^2 \over \sigma^2 S_i} \right)\left(\left< U \right> - U_i \right) = {\partial \mathbb{S} \over \partial S_i \partial U_i} 
\eeq
When cross partial derivatives are not equal, the value of the function depends on its configurational history (e.g. whether we change $U_i$ or $S_i$ first). This system is thus conservative, and reversible, since the evolution of the entropy is not path-dependent. The utility of the representative consumer is also conservative (and therefore so is the average utility), which means that over any possible closed trajectory, the value of the utility of the representative agent $\overline{U}$ does not change:
\beq
{\partial \overline{U} \over \partial U_i \partial S_i} = {\partial P_i \over \partial S_i} = {P_i \over S_i} - {P_i^2 \over S_i} = {\partial \overline{U} \over \partial S_i \partial U_i} \Rightarrow \quad  \oint_{\vec{T}} d\overline{U} = 0.
\eeq
Where $\vec{T}$ is any trajectory in $S_i$ or $U_i$. 

However, these things change when we change the technology list of innovations. As opposed to the MNL, we can do this quasi-continuously by including new innovations with near-to-zero market shares, consistent with how innovations diffuse. If we consider a path $\vec{T}'$ in which we add an innovation to the list along the way, this becomes path-dependent, and not reversible. But furthermore, this system also becomes path-dependent for any level of fluctuations in $S_i$ or $U_i$. Thus, unless the system is considered perfectly deterministic, it remains conservative and in equilibrium; but vanishingly small fluctuations break this equilibrium rather quickly, which we show next. 

We consider changes taking place in time, and decide to to create a path of utility that goes back on itself, such that the start and end points have the same utility value set $U_i$ (e.g. with tax policy that is introduced, and later withdrawn). This should not give a change of $\overline{U}$ at the final (start) point, as the shares progress but afterwards undo themselves back. Choosing any trajectory $\vec{T}$ in time to carry out a line integral,
\beq
0 = \oint_{\vec{T}} \left[ \sum_i {\partial \overline{U} \over \partial S_i}\bigg|_{U_i} {d S_i \over dt} + {\partial \overline{U} \over \partial U_i}\bigg|_{S_i} {d U_i \over dt} \right] dt
\eeq
which, using the fact that $\partial \overline{U} / \partial S_i = P_i / S_i$, while $\partial \overline{U} / \partial U_i = P_i$, can be written as 
\beq
 = \sigma \sum_i \oint_{\vec{T}} \left[ {P_i \over S_i} {dS_i \over dt} + P_i {dU_i \over dt}   \right] dt, 
\eeq
\beq
= \sigma \oint_{\vec{T}} \left( \left< {1 \over S_i} {dS_i \over dt} \right> + {1 \over \sigma} \left<{dU_i \over dt} \right> \right) dt = 0,
\eeq
i.e. the preferences weighted average relative changes of shares and change in utility in time. As the system turns back towards its starting point, these reverse their progress,\footnote{Imagine the impact of a step change in the utility of one of the products (e.g. due to a new tax). The shares will evolve away from that product gradually. Reversing history means a step change back to the original utility value. The shares will follow the same path backwards.} and both terms are zero, and this appears to be reversible.

Including small random events, we consider that the adoption $S_i$ and the utilities vary with stochastic noise terms $\epsilon^S_i$ and $\epsilon^U_i$.\footnote{To maintain $\sum_i S_i = 1$, $\epsilon^S_i$ must sum to zero.} This leaves two terms that do not cancel,
\beq
= \sigma \oint_{\vec{T}} \left( \left< {1 \over \epsilon^S_i} {d\epsilon^S_i \over dt} \right> + {1 \over \sigma} \left<{d\epsilon^U_i \over dt} \right> \right) dt \neq 0.
\eeq
Even for vanishingly small amplitude disturbances from random events, the system's evolution becomes path dependent due to the \emph{accumulation of random fluctuations}, which can readily be observed computationally.\footnote{E.g. it is observable on a computer due to rounding error on 16~bit floating point variables.} We name this `weak path-dependence', which leads to a typical `butterfly effect', where different simulations have outcomes that diverge from one another exponentially with simulation time span, due to different accumulated random fluctuations. This emerges strongly due to the non-linearity of the model.\footnote{Random fluctuations influence small populations significantly more than large populations, and as in e.g. stochastic biological population models, leads to non-zero probability of extinction at all times, but increasingly large as population sizes decrease relative to the magnitude of the fluctuations.} 

In the case of new products, we imagine that along the path, a new product is created by entrepreneurs and introduced in the market, at time $t_1$, with vanishingly small but non-zero shares. Assuming that it is an attractive product, shares will leak out towards this new product, that we denote $\ell$. The process before this happens is reversible, but after, it is not possible to `un-invent' product $\ell$, since as time passes, it may have growing shares, (i.e. not vanishingly small anymore), when reversing the context (e.g. removing a policy), the change in $\overline{U}$ does not go back to its starting value, leaving
\beq
= \sigma \oint_{\vec{T}} \left( \sum_{i \neq \ell} {1 \over S_i} {dS_i \over dt} + {1 \over S_\ell} {dS_\ell \over dt}   \right) dt \neq 0.
\eeq
We denote this `strong path-dependence'.

A key property of this system is that it implies increasing returns to adoption of innovations (i.e. adoptions leading to increased likelihood of adoptions). Increasing returns have been discussed extensively by \cite{Arthur1989, Arthur1987} and the same author in \cite{Anderson1989}. There, it is shown how the existence of increasing returns to adoption in a system of two products with identical benefits leads to spontaneous symmetry breaking: the dominance of a technology is determined by comparatively small historical events in early stages. In the presence of decreasing returns, small fluctuations cancel out and disappear in the aggregate, maintaining path-independence. In the presence of increasing returns, fluctuations cumulate, resulting in path dependence.

The introduction of an attractive innovation involves work (including in the thermodynamical sense): it represents an injection of potential utility that the system can henceforth reach. This allows the values of $\mathbb{S}, \overline{U}, \left< U \right>$ to reach a new range of values. This is irreversible, meaning that this new amount of utility cannot be taken out afterwards.

Strong path-dependence is radically different from weak path-dependence: the system takes a new trajectory, enabling, in effect, to explore areas of utility inaccessible in the past. Intuitively, one expects that only innovations providing users with higher utility are likely to successfully diffuse in the market. With a flow of innovations, the system can feature indefinitely increasing utility of the representative agent, as new products with higher utility are gradually adopted, and products with lowest utility are gradually phased out and forgotten. This, in effect, corresponds to Schumpeter's process of economic development, as we discuss in section~\ref{sect:V}. Innovations can also fail, and policy can also play a role in changing the state of the system.

\subsection{Theory for weakly interacting consumers in economics \label{sect:IV}}

It is now possible to build micro-foundations for a non-equilibrium consumer theory. We start from the random utility function of \emph{individual agents} who are interacting with their peers, and derive the macro (collective) behaviour. Having interactions means that \emph{consumers value information gained from other consumers}, and we can flesh out the interaction term of eq.~\ref{eq:IntLRUM}:
\beq
U_{i} = \beta^1_i V^1_i + \beta^2_i V^2_i + \beta^3_iV^3_i + ... + \alpha \log \left[ {1\over N} \sum_k^N \delta_{ki}\right] + ... + \epsilon_i,
\label{eq:randUtk}
\eeq
\beq
\delta_{ki} = \left\{ \begin{array}{ll}  1 \quad \text{if agent}\,k\,\text{owns or uses product}\, i  \nn\\
0 \quad \text{otherwise} \end{array}. \right.
\eeq
Here, $\log \delta_i(k)$ gives $-\infty$ amounts of utility for an unknown option $i$, and zero otherwise, reflecting that agents cannot choose what they do not know, and $N$ is the number of agents. $\alpha$ is a scaling parameter that we discuss below, in units of utility. If agent $k$ knows of option $i$, then the random utility has the same value as in eq.~\ref{eq:randUt}.

Agent $k$ will learn about option $i$ only if he can meet agent $\ell$ who has experience of option $i$,\footnote{See \cite{Durlauf2010}, in which the same form of utility function is used.} 
\beq
U_i = \beta^1_i V^1_i + \beta^2_i V^2_i + \beta^3_iV^3_i + ... + \alpha \log \left[ {2\over N N_p} \sum_k^N \sum_\ell^{M} \xi_{k \ell}\delta_{\ell i}\right] + ... + \epsilon_i,
\label{eq:randUtkell}
\eeq
\beq
\xi_{k \ell} = \left\{ \begin{array}{ll}  1 \quad \text{if agent}\,k\,\text{knows/trusts agent}\, \ell  \nn\\
0 \quad \text{otherwise} \end{array}, \right.  \quad M \equiv \text{Nearest neighbours.}
\eeq
Here the sum over $M$ is carried out over each agent's set of accessible peers, and $N_p$ is the average number of accessible peers per agent.\footnote{The average of M over agents equals $N_p$. The factor 2 avoids double counting pairs of agents. We assume that in a pair, interactions are bi-directional.}

Denoting $U_i^0$ as the random utility of non-interacting consumers for product $i$, inserting eq.~\ref{eq:randUtkell} into the utility of the representative agent eq.~\ref{eq:repU}, we obtain
\beq
\overline{U} = \sigma \log \left( \sum_i \exp \left[ {U_i^0 \over \sigma} + {\alpha \over \sigma} \log \left( {2\over N N_p} \sum_k^N \sum_\ell^{M} \xi_{k \ell}\delta_{\ell i}\right)  \right]   \right)
\eeq
Every pair of agents $k$ and $\ell$ where agent $\ell$ has no knowledge of product $i$, we have a contribution of $-\infty$ to the utility, which gives a contribution of zero to the utility of the representative agent. We can calculate further:
\beq
\overline{U} = \sigma \log  \left( \sum_i \left[ \sum_k^N \sum_\ell^{M} {2 \xi_{k \ell}\delta_{\ell i} \over N N_p} \right]^{\alpha \over \sigma} \exp\left({U_i^0 \over \sigma} \right) \right).
\eeq
We approximate that agents have on average $N_p$ trusted peers, and the sum over agents $k$ yields the number $N_{i}$ of agents that use product $i$. Noting that $N_i/N$ equals the share of product use $S_i$, we obtain
\beq
\overline{U} = \sigma \log  \left( \sum_i \left[ S_i \right]^{\alpha \over \sigma} \exp\left({U_i^0 \over \sigma} \right) \right).
\label{eq:repUintW}
\eeq
When the ratio $\alpha / \sigma$ is equal to 1, we recover the utility of the representative consumer for strongly interacting agents, eq.~\ref{eq:repU}. When the ratio is equal to zero, we recover that of non-interacting agents.

Eq.~\ref{eq:repUintW} can be differentiated in order to obtain preferences $P_i$:
\beq
P_i = {\partial \overline{U}\over \partial U_i} = {\left[ S_i \right]^{\alpha \over \sigma} \exp\left({U_i^0 \over \sigma} \right) \over \sum_{k} \left[ S_k \right]^{\alpha \over \sigma} \exp\left({U_k^0 \over \sigma} \right)}  
\label{eq:PrefStr}
\eeq
The shares equation would then be similar to \ref{eq:shares1}:
\beq
{dS_i \over dt} = {1 \over \tau_i}\left[ {\overline{\tau} \over \tilde{\tau}}{\left[ S_i \right]^{\alpha \over \sigma} \exp\left({U_i^0 \over \sigma} \right) \over \sum_{k} \left[ S_k \right]^{\alpha \over \sigma} \exp\left({U_k^0 \over \sigma} \right)}   - S_i \right]
\eeq
which, for durable products, can also be expressed as
\beq
{d S_i \over dt} = \sum_j S_i^{\alpha \over \sigma} S_j^{\alpha \over \sigma} \left( A_{ij}F_{ij} - A_{ji}F_{ji} \right), 
\eeq
We already know that if we tune $\tau_i$ towards zero, we obtain an equilibrium model. We now also find that if, in addition to tuning $\tau_i$ to zero, we also tune the multi-agent interactions $\alpha$ towards zero, we recover the MNL, where
\beq
S_i = {\exp\left({U_i^0 \over \sigma} \right) \over \sum_{k} \exp\left({U_k^0 \over \sigma} \right)} 
\eeq

Adding learning interactions between agents and time lags in consumption in this model therefore takes us from the MNL in eq.~\ref{eq:partial}, of standard consumer theory, to the replicator dynamics eq.~\ref{eq:Sprefs}, of evolutionary theory. The parameter $\alpha / \sigma$ can be used to alter the magnitude of multi-agent interactions. When $\alpha / \sigma \rightarrow 0$, , we \emph{`tune' the MNL in and out of equilibrium}. If both the price context changes much more slowly than the product purchase turnover rate $1/\tau_i$, and interactions are tuned to zero, standard consumer theory is recovered, the MNL, i.e. utility optimisation under constraint. 

This result also means that this problem is one of correlations between agents and their behaviour, a complex `many-body problem', where interactions in this case imply a `frustrated' dynamical system without equilibrium. As soon as we have learning interactions between agents with time delays, however small, the model becomes non-linear and goes out of equilibrium, and complexity, hysteresis and path-dependence arise.

\section{The supply of new products: business, innovation and diffusion} \label{sect:V}

\subsection{Innovation and the supply of new products}

\begin{figure}[t]
	\begin{center}
		\includegraphics[width=1\columnwidth]{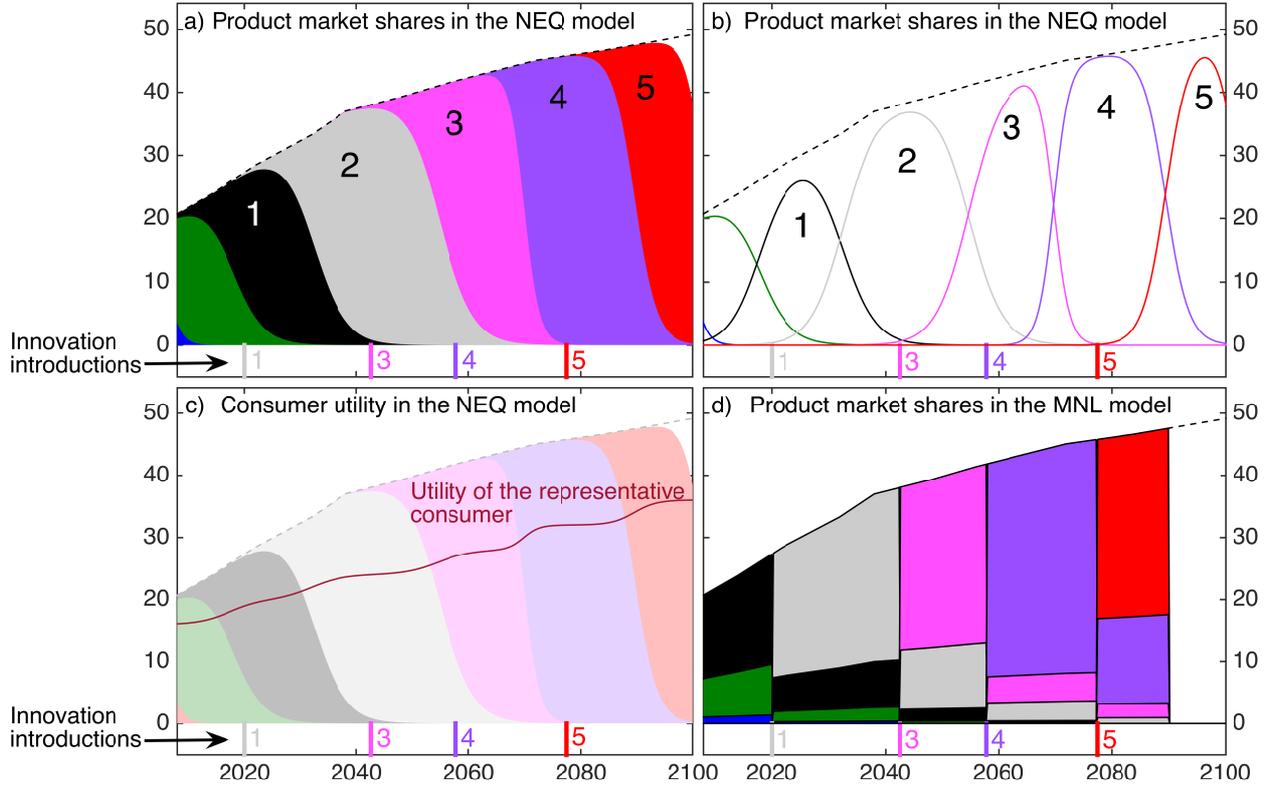}
	\end{center}
	\caption{Illustration of the non-equilibrium (NEQ) and multinomial logit (MNL) models, in a case where 5 innovative variants of a product are introduced at particular points in time (numbers 1-5 at bottom of the charts). \emph{a)-b)} Diffusion of these successive generations of innovations. The dotted line is the total number of units. \emph{c)} The utility of the representative agent increases indefinitely. \emph{d)} Outcomes of the MNL in a similar context: it discontinuously changes each time an innovation is introduced.}
	\label{fig:Figure2}
\end{figure} 

In our model, for any degree of social influence, however small, if the context remained static for a sufficiently long time, shares $S_i$ would catch up with preferences $P_i$ where either all utilities $U_i$ are equal (unstable equilibrium) or one product dominates the market (stable equilibrium). In the latter case, the model would also reach nil product diversity, inconsistent with observed reality. Product diversity does not mean maintaining the same products, as new/old products continuously enter/exit the market. This in/out flow provides ever higher utility to users, and not a steady state.\footnote{For example, the rate of innovation can change over time, and indeed changes all the time, for example if entrepreneurs in a particular area run out of ideas, or if a breakthrough happens. See for instance \cite{Arthur2006}.}

In Post-Schumpeterian theory \citep[e.g.][]{Safarzynska2010,Perez2001}, entrepreneurs create new products motivated by the prospect of securing monopoly rents. In the present model, this means a stream of products offering ever higher amounts of utility (through increased functionality, lower cost) which allows the utility of the representative agent to increase as long as the flow of innovations does not cease, in an ever evolving market.\footnote{For example, consider the mobile phone market, which continuously evolves, every generation offering higher numbers of possible applications, gradually transforming the way communication is done by users.} 

As we have shown, this process is irreversible and perpetually out of equilibrium. The replicator dynamics expresses the \emph{selection} process filtering innovations generated by entrepreneurs. The popular ones take off, the unpopular ones fail. This is illustrated in the top  panels of figure~\ref{fig:Figure2} (panels a-b). The succession of innovations generates ever increasing utility for the consumer (panel c). The source of irreversibility, which disrupts the equilibrium, is investment in R\&D leading to the introduction of new products (as well as fluctuations in adoption or utility). In this model, this implies continuously expanding the set $\{i\}$ of existing products. 

Three important aspects of this are to be noted. (1) The introduction of new products at vanishingly small shares $S_i$ does not discontinuously disturb the system even though it generates irreversibility. (2) A flow arises away from old products towards new ones that replace them. (3) The utility of the representative consumer indefinitely increases, and thus cannot be maximised in an optimisation algorithm.

The behaviour of the MNL (and CES models) is radically different to this when faced with innovation. The allocation of consumer choices changes if either utilities change, or if the list of products change. Where new products are introduced, the allocation of market shares changes discontinuously every time a new product is introduced (panel d). This is due to information or product access reaching consumers instantaneously, something contradicted by large amounts of empirical literature.\footnote{See for example \cite{Fisher1971, Bass1969, Rogers2010, Mansfield1961}.} In practice, equilibrium models are not typically used with changing lists of products.

\subsection{Replicator dynamical equations can augment CES functions in quantitative models for the substitution process}

Economic modellers do not always grasp the full meaning of their assumptions when they use optimisation algorithms. Here, we have effectively demonstrated that using a MNL or CES framework excludes the process of diffusion of innovations as promoted by social interactions. Indeed, CES utility functions (or the variants Stone-Geary, Cobb-Douglas) are used in most General Equilibrium models, as they offer a convenient analytical form for optimisation. 

To include innovation diffusion in an economic model, the CES could in principle be replaced by the replicator dynamics (or Lotka-Volterra) equation, which then requires using a time step simulation (a dynamical system). The latter can also be formulated in a nested form\footnote{I.e. a decision tree with choices and sub-choices, etc, as with nested MNLs or nested CES), in the same way with the same benefits. Note that as in statistical mechanics, nesting a MNL or CES model serves the purpose of organising the degrees of freedom of the decision problem. It gives rise to a multiplicative concatenation of partition functions. The same can be said of the replicator dynamics.} This means that whole optimisation models (partial equilibrium, CGE) could in principle be transformed into simulations using the same databases.

The most convenient form to use computationally is the Lotka-Volterra set of differential equations:
\beq
{d S_i \over dt} = \sum_j S_i^{\alpha \over \sigma} S_j^{\alpha \over \sigma} \left( A_{ij}F_{ij} - A_{ji}F_{ji} \right), 
\eeq
One requires estimates of the strength of the interaction parameter $\alpha/\sigma$ and sector-dependent turnover rates $\tau$. Consumer preference distributions can be inferred from cross-sectional data.\footnote{As in \cite{MercureLam2015}, it is possible to parameterise distributions cross-sectionally using costs instead of utilities} Where social influence is not important (e.g. commodity trade), the MNL can be used. Note, however, that determining $\alpha$ involves problems of identification \citep{Manski1993}.

\section{Conclusions} \label{sect:VI}


In standard consumer theory, given that choice options can be ranked, are transitive and known by all agents, the representative consumer is able to   exhaustively rank his preferences based on those the underlying population that he represents. We have shown that in a model of learning agents with heterogenous knowledge, the representative agent is not able to establish a ranking of his preferences, because he represents agents each of which establishes his/her preferences on a different subset of options, and none of which know all products in the market. Every agent maximises his utility on a different subset of products that he knows. However, preferences of agents `attract' the preferences of other agents through interactions (social influence), and therefore preferences cluster in groups (social groups). In a model with learning agents and heterogenous knowledge, there does not exist any `objective' set of preferences shared by everyone, and no product or social choice is `objectively' superior than any other, since consensus (a majority preference) on any cannot be obtained across all agents simulataneously. 

The deeper meaning of this finding is that in a theory of learning agents with heterogenous knowledge, preferences and choices are subjective to context and whom it refers to: social identity groups, income classes, cultural origins, professional classes, etc. The meaning of utility is itself subjective to every agent and his context, and not comparable, aggregable or averageable across agents. This model of learning heterogenous agents has a number of implications:

First, the representative agent loses meaning as soon as infinitesimally small interactions (social influence) occur between agents. With non-zero interactions and a rapidly changing context, the multinomial logit becomes the the non-equilibrium replicator dynamics system given in this paper. That is, a dynamical evolutionary model of selection and diffusion. It also follows that standard consumer models (MNL, elasticity of substitution models) are only adequate for systems of isolated agents without interactions in a relatively static context (the market settles much faster than changes in context or the rate of innovations). 

Second, optimisations and Lagrangians are not reliable analysis tools for a models of learning agents. Such a model is a dynamical system in perpetual motion without equilibrium, but with hysteresis. The timescale for equilibrium is longer than the measurement timescale, and longer than the average time between innovations introduced, each of which push the system further out of equilibrium. 

Third, the utility of the representative agent thus \emph{cannot reliably be used as a measure of social welfare}, since 
\begin{enumerate}
\item \emph{It does not represent everyone equally}. At any time, some agents gain more utility from contextual changes (e.g. taxes) than others.
\item \emph{Its meaning is ambiguous}. Its relationship with the utility of any particular individual agent is not straightforward to determine.
\item \emph{It cannot be maximised (its maximum cannot be found)}. With increasing returns, fads and fashions appear, and utility is unbounded and path-dependent. Path-dependence makes the state of the economy subjective to its context and configurational history. 
\end{enumerate}

Fourth, the purpose of the act of consumption is clearer in a model of learning heterogenous agents, which enables linking to standard socio-anthropological theory. In socio-anthropological theory, consumption is context-dependent \citep{Douglas1979}. This is for example expressed as different consumption habits held by different income, social, cultural or professional groups. Particular cultural groups consume particular types of food products much more rooted in their tradition, in order to follow their kin, rather than based on their relative price (in comparison to what other groups consume) and the absolute utility (e.g. nutrition) they derive from it. In other words, the utility associated with tradition (behaving like peers) is at least as important as the utility associated with the product itself. 

Fifth, in standard theory, it is typically difficult to explain the simultaneous observations of (1) relative product diversity across social groups and (2) relative product homogeneity within social groups \citep{Young2001}. It is tenuous with the representative agent to explain why prices can differ markedly across markets serving different subgroups, for products serving near identical purposes \citep[e.g. cars, where prices vary by orders of magnitude, see][]{MercureLam2015}. It is much more plausible that different social groups choose within submarkets for goods that perform similar functions, and  do not delve very strongly in each other's submarkets \citep[e.g.][]{McShane2012}. The present model is consistent with these observations.

Sixth, while in standard models, optimal tax policy determines the quantities of goods consumed in equilibrium, in a non-equilibrium model, tax policy influences the trajectory of diffusion, but not directly the quantity of goods sold. 

\appendix


\section{Appendix A: Deriving the MNL from the CES model} \label{sect:AppA}

We have the CES utility function $U$ and budget constraint $Y$
\beq
U = \left( \sum_j \left[X_j\right]^\rho \right)^{1\over \rho}, \quad Y = \sum_j X_i p_i,
\eeq
where $X_i$ are quantities of goods and $p_i$ are prices. We want to maximise $U$ under constraint $Y$. We define the Lagrangian $L = U - \lambda Y$ and take gradients over all parameters $X_i$:
\beq
\nabla_{X_i} L = 0 \Rightarrow \nabla_{X_i} U =  \lambda \nabla_{X_i} Y,
\eeq
from which we derive
\beq
\left( \sum_j \left[X_j \right]^\rho \right)^{1-\rho \over \rho} X_i^{\rho-1} = \lambda p_i.
\eeq
and obtain
\beq
{X_i \over \left( \sum_j X_j^{\rho}\right)^{1 \over \rho}} = \left( \lambda p_i \right)^{1 \over \rho - 1}
\eeq
We invert this relationship by substituting $X_j$ in the denominator for the value of $X_i$ itself:
\beq
X_j = \left(\lambda p_j\right)^{1 \over \rho - 1} \left( \sum_k X_k^\rho \right)^{1\over \rho} = \left(\lambda p_j\right)^{1 \over \rho - 1} U,
\eeq
and obtain
\beq
\left( \lambda p_i \right)^{1 \over \rho - 1} = { X_i \over \left( \sum_j \left( \lambda p_j \right)^{\rho \over \rho - 1} U^{\rho} \right)^{1 \over \rho}}
\eeq
$X_i$ can be isolated,
\beq
X_i = \left( \lambda p_i \right)^{1 \over \rho - 1}  \left( \sum_j \left( \lambda p_j \right)^{ \rho \over \rho - 1}\right)^{1\over \rho} U
\eeq
If we define choices $P_i$ as shares of expenditure, in which the term $\left( \sum_j \left( \lambda p_j \right)^{ \rho \over \rho - 1}\right)^{1\over \rho} U$ cancels out, then:
\beq
P_i = {p_i X_i \over \sum_j p_j X_j} = {\left( \lambda p_i \right)^{\rho \over \rho - 1} \over \sum_j \left( \lambda p_j \right)^{\rho \over \rho - 1}} = {e^{U_i \over \sigma} \over \sum_i e^{U_j \over \sigma}},
\eeq
i.e. equal to the MNL. Taking the utility $U_i$ as a logarithmic function of prices $-\log \lambda p_i$ implies that
\beq
{U_i \over \sigma} = {\rho \over 1 - \rho} \log \lambda p_i, \quad \text{with}  \quad 0<\rho<1.
\eeq
This determines the price and cross-price elasticities. This shows a complete equivalence between the MNL and the CES. Further information can be obtained in \cite{Anderson1992}, section 3.7.

\section{Appendix B: Demonstrating non-equilibrium dynamics} \label{sect:AppB}

We prove this point by contradiction. We make a fairly general assumption about the form of $P_i$ by expressing it in terms of the Fourier transforms of the numerator and denominator:
\beq
P_i = { \sum_n A_{ni} e^{\alpha_n {U_i \over \sigma_n}}\over \sum_{mk} A_{mk} e^{\alpha_m {U_k \over \sigma_m}}},
\label{eq:Pi}
\eeq
where the $A_{ni}$ are real Fourier coefficients and $\alpha_n = i u_n + v_n$ are complex frequencies. We need the derivative of this term:
\beq
{\partial P_j \over \partial U_i} = 
{ \sum_{nj} {\alpha_n \over \sigma }A_{nj} e^{\alpha_n {U_j \over \sigma_n}}\delta_{ij} \over \sum_{mk} A_{mk} e^{\alpha_m {U_k \over \sigma_m}} }
+ { \left(\sum_{nj} A_{nj} e^{\alpha_n {U_j \over \sigma_n}}\right)\left( \sum_{mj} {\alpha_m \over \sigma }A_{mj} e^{\alpha_m {U_j \over \sigma_m}} \delta_{ij} \right) \over \left( \sum_{\ell k} A_{\ell k} e^{\alpha_\ell {U_k \over \sigma_\ell}} \right)^2},
\label{eq:dPjdUi}
\eeq
where $\delta_{ij}$ is Kronecker's delta, equal to zero unless $i=j$, in which case it equals 1. \footnote{$\delta_{ij} = 1 if i = j,$ but zero otherwise.}

Inserting eqns~\ref{eq:Pi} and~\ref{eq:dPjdUi} into eq.~\ref{eq:nonsense}, we obtain, after cancelling terms in the numerators and denominators:
\beq
{ \sum_n A_{ni} e^{\alpha_n {U_i \over \sigma_n}}\over \sum_{mk} A_{mk} e^{\alpha_m {U_k \over \sigma_m}}} 
= { \sum_n A_{ni} e^{(\alpha_n+1) {U_i \over \sigma_n}}\over \sum_{mk} A_{mk} e^{(\alpha_m+1) {U_k \over \sigma_m}}},
+{ \sum_n \alpha_n A_{ni} e^{(\alpha_n+1) {U_i \over \sigma_n}}\over \sum_{mk} A_{mk} e^{(\alpha_m+1) {U_k \over \sigma_m}}},
-{ \sum_n \alpha_n A_{ni} e^{\alpha_n {U_i \over \sigma_n}}\over \sum_{mk} A_{mk} e^{\alpha_m {U_k \over \sigma_m}}}.
\eeq
This reduces to
\beq
{ \sum_n A_{ni} e^{\alpha_n {U_i \over \sigma_n}}\over \sum_{mk} A_{mk} e^{\alpha_m {U_k \over \sigma_m}}} (\alpha_n + 1)
= { \sum_n A_{ni} e^{(\alpha_n+1) {U_i \over \sigma_n}}\over \sum_{mk} A_{mk} e^{(\alpha_m+1) {U_k \over \sigma_m}}} (\alpha_n + 1),
\eeq
where the $(\alpha_n + 1)$ cancel on each side. We are then led to equating $\alpha_n = \alpha_n + 1$ for any complex set of values for $\alpha_n$, which is a contradiction. Thus this general form for $P_i$ does not solve eq.~\ref{eq:nonsense}, unless one of the $P_i$ equals one and the others equal zero, which is the trivial solution.

\section{Appendix C: Connecting different forms of the replicator dynamics} \label{sect:AppC}

Here we make the connection between a binary exchange model \citep[imitation dynamics, ][]{Hofbauer1998}, and the replicator dynamics obtained from adding social interactions in the MNL, eq.~\ref{eq:replicator}. In order to obtain a binary exchange model, we consider a system of boxes and marbles in fixed numbers. The system consists in distributing and redistributing marbles from box to box according to rates determined by the constants $t_i$ and $\tau_i$, one time constant limiting the maximum rate of growth, and the other limiting the fastest rate at which a box can be emptied. This leads to dynamics extensively described in \cite{Mercure2015}, a Lotka-Volterra equation system expressed in the form of shares of the total.

We consider the replicator dynamics of the classical form,
\beq
{d S_i \over dt} = {S_i \over \overline{\tau}} \left( \mathcal{F}_i - \overline{\mathcal{F}} \right), \quad \mathcal{F}_i = P_i - P_i'
\eeq
in which the fitness for survival in the market of product $i$ is expressed using $\mathcal{F}$, and $\overline{\mathcal{F}}$ is the average fitness, and the fitness is related to the probability of adoption $P_i$ and the probability of scrapping $P_i'$.  The probability of adoption is greater the greater the utility (without interactions) $U_i^0$, changes that arise at the average rate of turnover $\overline{\tau}$,
\beq
{d S_i^\uparrow \over dt} = P_i =  { {\overline{t} \over t_i} S_i \exp\left({U_i^0 \over \sigma} \right) \over \sum_{k} {\overline{t} \over t_k} S_k \exp\left({U_k^0 \over \sigma} \right)}.
\eeq
The economic decision to scrap products arises with a probability greater the lower the utility is,
\beq
{d S_i^\downarrow \over dt} = - P_i' =  - { {\overline{\tau} \over \tau_i} S_i \exp\left({-U_i^0 \over \sigma} \right) \over \sum_{k} {\overline{\tau} \over \tau_k} S_k \exp\left({-U_k^0 \over \sigma} \right)}.
\eeq
Then the total fitness for survival of the product in the market is
\beq
\mathcal{F}_i = { {\overline{t} \over t_i} S_i \exp \left({U_i^0 \over \sigma} \right) \over \sum_{k} {\overline{t} \over t_k} S_k \exp\left({U_k^0 \over \sigma} \right)} - { {\overline{\tau} \over \tau_i} S_i \exp\left({-U_i^0 \over \sigma} \right) \over \sum_{k} {\overline{\tau} \over \tau_k} S_k \exp\left({-U_k^0 \over \sigma} \right)} 
\label{eq:Bigdenom}
\eeq
\beq
= { \sum_j S_i S_j \left( {\overline{t} \over t_i} {\overline{\tau} \over \tau_j}e^{(U_i - U_j)/\sigma} - {\overline{t} \over t_j} {\overline{\tau} \over \tau_i} e^{(U_j - U_i)/\sigma}\right) \over \left(\sum_{k} S_k {\overline{t} \over t_k}  e^{U_k/\sigma}\right)\left(\sum_{\ell} S_\ell {\overline{\tau} \over \tau_\ell} e^{ - U_\ell/\sigma}\right) },\nn
\eeq
while the average fitness equals zero in this form. The denominator can be distributed as
\beq
\left(\sum_{k} S_k {\overline{t} \over t_k}  e^{U_k/\sigma}\right)\left(\sum_{\ell} S_\ell {\overline{\tau} \over \tau_\ell} e^{ - U_\ell/\sigma}\right) = \sum_{k, \ell} {\overline{t} \over t_k} {\overline{\tau} \over \tau_\ell} S_k S_\ell e^{(U_k - U_\ell)/\sigma}
\eeq
and shown to be approximately constant, and reduced by symmetry, for every pair-wise $k,\ell$ case where the difference in utility is not large: assuming that $\tau_\ell$ is not quantitatively very different than the average $\overline{\tau}$, and similarly for $t_k$, then it becomes approximately
\beq
\simeq \sum_{k, \ell} S_k S_\ell e^{(U_k - U_\ell)/\sigma} = {1\over 2} \sum_{k, \ell} S_k S_\ell e^{(U_k - U_\ell)/\sigma} + {1\over 2} \sum_{\ell, k} S_\ell S_k e^{(U_\ell - U_k)/\sigma} 
\eeq
\beq
= \sum_{k, \ell} S_k S_\ell \cosh(U_k - U_\ell)/\sigma.
\eeq
If the value of $U_k - U_\ell$ is not much larger than that of $\sigma$, then $\cosh (U_k - U_\ell)/\sigma \simeq 1 +(U_k - U_\ell)^2/\sigma^2$, meaning that the denominator reduces to something very close to a value of 1. This will be the case when the sum of the Gumbel distributions of the alternatives has a single mode, in which case the weighted average of the distributions is not highly different from the distribution of the representative consumer. 

However, whenever one option has a utility very different from the others ($U_k >> U_\ell$), this will not be the case, as the exponential will have either a very large or near zero value (depending on the sign). In this case, we carry out the sum over $k$ over only those cases $i$ and replace the alternate $\ell$ utility by the average utility $\overline{U}$, (i.e. $U_\ell \simeq \overline{U}$):
\beq
\simeq S_i \sum_\ell S_\ell \left( {\overline{t} \over t_i} {\overline{\tau} \over \tau_\ell} e^{(U_i - \overline{U})/\sigma} + {\overline{t} \over t_\ell} {\overline{\tau} \over \tau_i} e^{(\overline{U} - U_i)/\sigma} \right) \simeq S_i \left( e^{(U_i - \overline{U})/\sigma} + e^{(\overline{U} - U_i)/\sigma} \right).
\eeq
In this case, the denominator of eq.~\ref{eq:Bigdenom} can be integrated to the numerator. If options $i$ and $j$ differ in utility significantly, and option $i$ is identified as the option differing from the average, while $j$ remains similar to the average, then the denominator, merged to the numerator, gives logistic functions of $U_i - U_j / \sigma$ (by identifying $U_j$ to $\overline{U}$). If $i$ and $j$ do not differ much, the denominator remains close to 1 and the numerator close to 0, and thus this approximation remains correct. 

Under these assumptions, the replicator dynamics becomes approximately
\beq
{d S_i \over dt} = {1\over \overline{\tau} }\sum_j S_i S_j \left( A_{ij} F_{ij} - A_{ji}F_{ji}\right),
\eeq
\beq
\text{where} \quad F_{ij} = {e^{(U_i - U_j)/ \sigma} \over e^{(U_i - U_j)/ \sigma} + e^{(U_j - U_i)/ \sigma}}, \quad F_{ij} + F_{ji} = 1, \quad A_{ij} = {\overline{t} \over t_i} {\overline{\tau} \over \tau_j}.\nn
\eeq
This is a pair-wise model of choice and substitution, analogous to a Lotka-Volterra model which can be derived under pair-wise interaction assumptions \citep[as in the population dynamics of interacting species, see][]{Mercure2015}, also called `imitation dynamics' in \cite{Hofbauer1998}. It can be seen as exchanges of market shares $S_i$ arising between pairs of options according to their utility differences. 

If more than one option differs from the average, and in a way that is different, this approximation loses accuracy over the exchange between the items of the pair $i,j$ for which both utilities differ from the average. In this case, the pair-wise model and the replicator dynamics equations give different results. It may be argued that the pair-wise interaction model more accurately represents reality than a model where a comparison is made to the average fitness \citep{Mercure2015}; however, the difference is not large. 

Both models are stock-flow models of product accumulation and depreciation, or in other words, a vintage capital model, useful for situation such as vehicle fleets, or industry plant fleets, in which the modeller wishes to keep track of ageing capital and its probability of surviving.

\section*{References}

\bibliographystyle{elsarticle-harv}
\bibliography{../../CamRefs}

\end{document}